\documentclass{article}
\usepackage[T1]{fontenc}
\usepackage[utf8]{inputenc}
\usepackage{ismir} 
\usepackage{amsmath,cite,url}
\usepackage{graphicx}
\usepackage{color}
\usepackage{booktabs}
\usepackage{placeins}
\usepackage{amssymb}
\usepackage{pifont}

\usepackage{algorithm} 
\usepackage{algpseudocode} 
\usepackage{gensymb} 
\usepackage{siunitx} 
\usepackage{multirow} 
\usepackage{siunitx} 
\usepackage{kotex}
\newcommand{\cmark}{\ding{51}}%
\newcommand{\xmark}{\ding{55}}%
\usepackage{microtype} 
\usepackage{paralist} 
\usepackage{float} 

\title{PianoVAM: A Multimodal Piano Performance Dataset}

\multauthor
  {Yonghyun Kim$^\flat$ \hspace{1cm} Junhyung Park$^\natural$ \hspace{1cm} Joonhyung Bae$^\sharp$ \hspace{1cm} Kirak Kim$^\sharp$}
  {{\bf Taegyun Kwon$^\sharp$ \hspace{1cm} Alexander Lerch$^\flat$ \hspace{1cm} Juhan Nam$^\sharp$}\\
  $^\flat$ Music Informatics Group, Georgia Institute of Technology, USA\\
  $^\natural$ Department of Mathematical Sciences, KAIST, South Korea\\
  $^\sharp$ Graduate School of Culture Technology, KAIST, South Korea\\
  {\small \{yonghyun.kim, alexander.lerch\}@gatech.edu,
  \{tonyishappy, jh.bae, kirak, ilcobo2, juhan.nam\}@kaist.ac.kr}
  }

\def\authorname{Y.~Kim, J.~Park, J.~Bae, T.~Kwon, K.~Kim, A.~Lerch and J.~Nam}

\usepackage[bookmarks=false,pdfauthor={\authorname},pdfsubject={\pdfsubject},hidelinks]{hyperref}

\begin{document}

\maketitle

\begin{abstract}
The multimodal nature of music performance has driven increasing interest in data beyond the audio domain within the music information retrieval (MIR) community. This paper introduces PianoVAM, a comprehensive piano performance dataset that includes videos, audio, MIDI, hand landmarks, fingering labels, and rich metadata. The dataset was recorded using a Disklavier piano, capturing audio and MIDI from amateur pianists during their daily practice sessions, alongside synchronized top-view videos in realistic and varied performance conditions. Hand landmarks and fingering labels were extracted using a pretrained hand pose estimation model and a semi-automated fingering annotation algorithm. We discuss the challenges encountered during data collection and the alignment process across different modalities. Additionally, we describe our fingering annotation method based on hand landmarks extracted from videos. Finally, we present benchmarking results for both audio-only and audio-visual piano transcription using the PianoVAM dataset and discuss additional potential applications.

\end{abstract}

\begin{figure}
    \centering
    \includegraphics[width=1\linewidth]{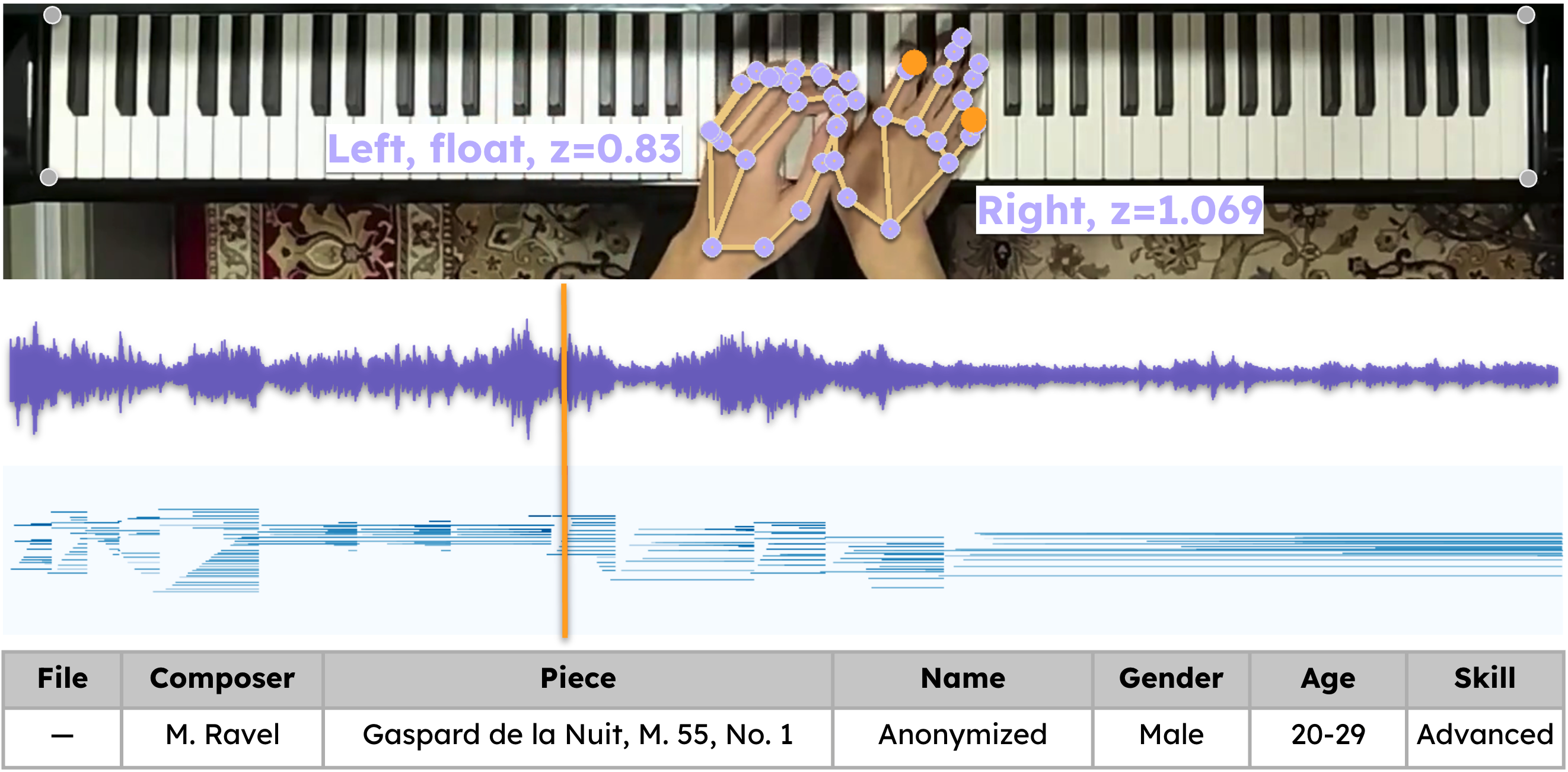}
    \caption{The PianoVAM Dataset: Synchronized video, audio, MIDI with fingering, hand landmarks, and metadata.}
    \label{fig:overview}
\vspace{-5mm}    
\end{figure}

\section{Introduction}\label{sec:introduction} 
Music performance is inherently multimodal, involving not only audio but also motion, posture, and other visual elements as part of the expressive sound creation process \cite{PPR09Bergeron, MusicPercep12Platz}. In the field of Music Information Retrieval (MIR), there has been growing interest in collecting multimodal performance data to enhance the extraction of musical information by leveraging multiple modalities \cite{TMM18Li, IEEE19Duan}. Such multimodal data ---particularly audio-visual data--- have been utilized in various MIR tasks across diverse musical genres. Examples include automatic MIDI transcription of solo piano performances \cite{ICASSP20Koepke, ICASSPW23Li}, vibrato analysis of polyphonic string music \cite{SMC17Li}, singing voice separation \cite{BMVC21Montesinos}, and melodic motif identification in Indian vocal performances \cite{TISMIR24Nadkarni}. In these tasks, visual information from performer videos has been shown to improve model robustness by providing additional musical cues. This paper focuses on the multimodal data collection of solo piano performances to improve transcription and explore other potential applications.

\begin{table*}[t]
    \centering
    \small
    \begin{tabular}{lcccccc}
        \toprule
        \textbf{Dataset}  & \textbf{Size (hrs)} & \textbf{Video} & \textbf{Audio} & \textbf{Audio Type} & \textbf{MIDI} & \textbf{Fingering} \\
        \midrule
        MAESTROv3 \cite{ICLR19Hawthorne}  & 198.7  & \xmark  & {44.1--48}\si{kHz}, Stereo & Real  & \cmark & \xmark \\
        MAPS (MUS subset) \cite{Emiya2010}     & 18.6  & \xmark     & 44.1\si{kHz}, Stereo & Synth. \& Real & \cmark & \xmark \\
        OMAPS2 \cite{ICASSPW23Li}   & 6.7  & 1080p/30fps & 44.1\si{kHz}, Mono & Real & $\triangle$ (.TXT)  & \xmark \\
        PianoYT \cite{ICASSP20Koepke}  & $\sim$20  & Varies  & Varies (YouTube) & Varies (YouTube) & $\triangle$ (Pseudo) & \xmark \\
        PianoVAM (Ours) & 21.0 & 1080p/60fps & 44.1\si{kHz}, Mono & Real  & \cmark & $\triangle$ (Pseudo) \\
        \bottomrule
    \end{tabular}
\vspace{-2mm}    
    \label{tab:piano_datasets}
\caption{Comparison of piano transcription datasets.}
\end{table*}

\begin{table*}
    \centering

    \small
    \begin{tabular}{lcccccc}
        \toprule
        \textbf{Dataset}  & \textbf{Total notes} & \textbf{\# of pieces} & \textbf{Labeled ratio (\%)} & \textbf{Data type} & \textbf{Reliability} & \textbf{Annotation} \\
        \midrule
        PIG \cite{InfoSci20Nakamura} & 100,044 & 309 & 100 & MIDI \& Score (.PDF) & By pianists & Manual \\
        ThumbSet \cite{MM22Ramoneda} & -- & 2,523 & 52 & MusicXML & By miscellaneous & Manual \\
        PianoVAM (Ours) & 1,050,966 & 106 & 100 & Multimodal & $\sim$ 0.95 & Semi-Auto \\
        \bottomrule
    \end{tabular}
\vspace{-2mm}    
\caption{Comparison of piano fingering datasets.}
    \label{tab:fingering-datasets}
\vspace{-2mm}    
\end{table*}

Piano transcription, which converts audio recordings into symbolic representations like MIDI or sheet music, is a well-established MIR task that has made significant progress through large-scale, clean audio-MIDI datasets such as MAESTRO \cite{ISMIR18Hawthorne} and carefully designed deep learning models \cite{TASLP21Kong, ISMIR22Wei}. As benchmarking performance on the MAESTRO dataset approaches its ceiling \cite{ISMIR24Yan}, new challenges have emerged in piano transcription. One major challenge is achieving acoustic robustness to ensure reliable transcription from real-world piano performance recordings, which often feature diverse piano timbres, reverberation, or other interfering noise sources. While data augmentation has been a common technique to address this issue \cite{ICLR19Hawthorne, Edwards2024}, leveraging visual data from performance videos has recently emerged as an alternative research direction \cite{CJE15Wan, DAFx21Wang, ICASSPW23Li, TASLP24Li}. Other key challenges include capturing richer performance information beyond a single MIDI track, such as left-right hand separation, piano fingering, and other playing details \cite{InfoSci20Nakamura, MM22Ramoneda}. Addressing these challenges requires capturing visual-domain data, such as performer motion or keyboard-view videos, and synchronizing them with audio and MIDI data. However, collecting such multimodal data is costly, requiring a dedicated data acquisition system, and time-consuming, as it depends on the availability of prepared piano players. 

This paper introduces PianoVAM, a comprehensive piano performance dataset that includes videos, audio, MIDI, hand landmarks, fingering labels, and rich metadata. An overview of PianoVAM is presented in Figure \ref{fig:overview}. The dataset was collected from amateur pianists during their daily practice sessions on a Disklavier piano, with synchronized top-view videos captured in realistic and varied performance conditions. We extracted hand landmarks and generated fingering pseudo-labels using a pretrained hand pose estimation model combined with a semi-automated fingering detection algorithm.  We describe the challenges faced during data collection and the alignment of multiple modalities. Furthermore, we detail our fingering annotation method, which utilizes hand landmarks extracted from the videos. Lastly, we present experimental results on both audio-only and audio-visual piano transcription using the PianoVAM dataset for benchmarking, along with a discussion of its potential applications. PianoVAM is available for download from the GitHub page\footnote{https://yonghyunk1m.github.io/PianoVAM\label{github-link}} under the CC BY-NC 4.0 license.

\section{Related Work}
\subsection{Audio-Visual Datasets}
The emergence of audio-visual datasets represents a promising frontier in MIR, unlocking new research possibilities by providing visual information that complements audio signals. The URMP dataset \cite{TMM18Li} offers synchronized audio, video, and MIDI recordings of multi-instrument classical performances, supporting multimodal analysis of ensemble music such as audio-visual source association via vibrato modeling \cite{SMC17Li}. The Acappella dataset \cite{BMVC21Montesinos} contains solo a cappella videos, enabling audio-visual singing voice separation with fine-grained control over visual and acoustic conditions. Nadkarni et al.\ present an audio-visual dataset of Hindustani vocal performances annotated with melodic motifs and stable notes, enabling gesture-based music analysis and raga classification through movement-melody correspondence \cite{TISMIR24Nadkarni}.

\subsection{Piano Transcription Datasets}
Table~\ref{tab:piano_datasets} compares existing piano performance datasets with PianoVAM. MAESTRO \cite{ICLR19Hawthorne} includes high-quality audio and MIDI data recorded from proficient pianists on Disklavier pianos but lacks top-view videos. While MAPS \cite{Emiya2010} provides audio and MIDI from actual performances (MUS subset), a significant portion (210 out of 270 recordings) is synthesized. OMAPS2 \cite{ICASSPW23Li} and PianoYT \cite{ICASSP20Koepke} incorporate video data but offer only limited MIDI annotations: OMAPS2 provides MIDI-like labels, while PianoYT uses pseudo-MIDI annotations transcribed with the Onsets and Frames model \cite{ISMIR18Hawthorne}. In comparison, PianoVAM offers the most comprehensive multimodal dataset, including real performance audio, synchronized MIDI, top-view videos, and fingering pseudo-labels, although its total duration is limited compared to the larger datasets listed.

\subsection{Fingering Datasets}
Table \ref{tab:fingering-datasets} compares existing piano fingering datasets with PianoVAM. PIG \cite{InfoSci20Nakamura} incorporates fingering and MIDI information of sections of several pieces played by professional pianists, which also provides different fingerings for the same piece by various pianists. Ramoneda et al.~\cite{MM22Ramoneda} attempted to annotate fingering of the complete piece from partially annotated MusicXML files with the support of ThumbSet dataset, which in turn crowd-sourced fingering information of numerous pieces from MuseScore\footnote{\href{https://musescore.com}{https://musescore.com} (Last accessed: September 10, 2025)}, an online piano score website, but the source of fingering annotation is not clear. The presented PianoVAM dataset, in contrast, utilizes a fingering detection algorithm  applied to top-view video data synced with MIDI and is improved by manual annotation of incomplete fingering labels to improve reliability.  

\section{Dataset Acquisition \& Pre-processing}\label{sec:dataset-acquisition-preprocessing}

\subsection{Acquisition}
We developed a data acquisition system to streamline the unsupervised recording of video, audio, MIDI, and associated metadata, such as performer and piece details.  
 
\begin{figure*}
    \centering
    \includegraphics[width=1\linewidth]{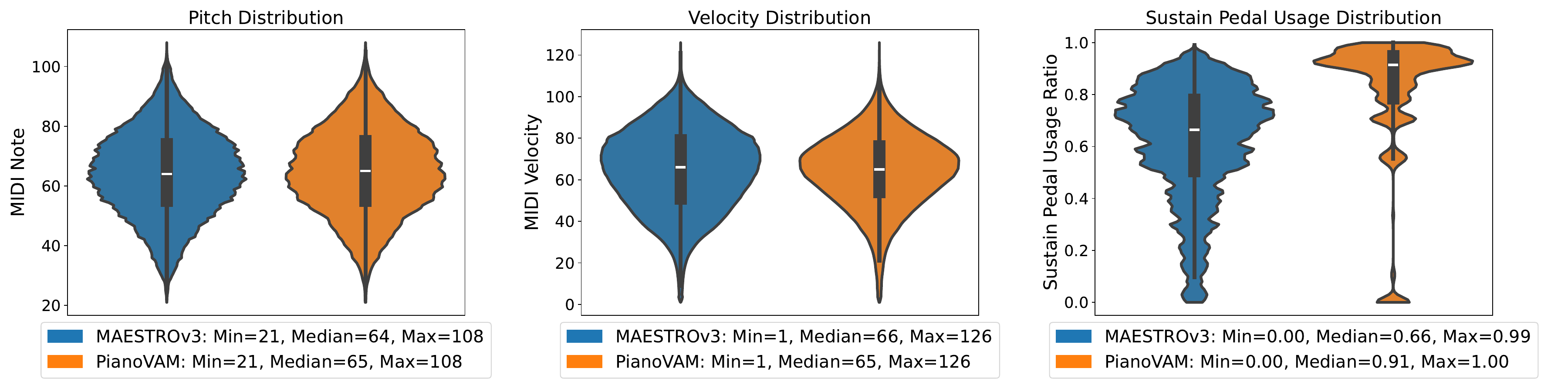}
    \vspace{-7mm}    
    \caption{Distributional comparison of pitch, velocity, and pedal usage between MAESTROv3 and PianoVAM.}
    \vspace{-3mm} 
    \label{fig:combined-violinplot}
\end{figure*}

\subsubsection{Acquisition Workflow}
The acquisition workflow comprises the following steps. First, new users register by providing basic personal details and receive unique QR codes for identification and recording control. A single interface launch initializes the required software: OBS Studio for video and audio, and Logic Pro for audio and MIDI. Before recording, users input performance details and display their profile QR code to a top-view camera to begin capturing. When the performance concludes, presenting a stop QR code ends the recording.

\subsubsection{Hardware Setup} 
The data acquisition process was managed by a control script on a PC. An overhead-mounted webcam captured the video, while a dedicated microphone and a Disklavier piano provided the audio and MIDI signals, respectively. OBS Studio recorded the audio-video stream, and Logic Pro concurrently recorded the audio-MIDI stream. A common audio signal captured by both systems provided a reference for the global time alignment to correct for recording latency.

\subsection{Pre-processing}
\subsubsection{Alignment}\label{subsubsec:alignment}
The time alignment of audio and MIDI data was further refined using the fine alignment technique used for the MAESTRO dataset \cite{ICLR19Hawthorne}. 
Specifically, the recorded audio was down-mixed to mono and resampled to {22.05}\si{kHz}. 
The MIDI data was then synthesized into an audio signal at the same sampling rate using \textit{FluidSynth} with a SoundFont sampled from Disklavier Pro recordings,\footnote{\href{https://freepats.zenvoid.org/Piano/YDP-GrandPiano/YDP-GrandPiano-SF2-20160804.tar.bz2}{https://freepats.zenvoid.org/Piano/YDP-GrandPiano/YDP-GrandPiano-SF2-20160804.tar.bz2} (Last accessed: September 10, 2025)\label{soundfont}} since we could not access the original Disklavier7 SoundFont.
A Constant-Q Transform was applied to  both audio signals using a hop length of 64 samples ($\sim$3\si{ms} resolution). Finally, we applied Dynamic Time Warping within a Sakoe-Chiba band of $\pm2.5$\si{s} to correct any remaining temporal discrepancies, such as small constant offsets or jitter.

\subsubsection{Audio Loudness Normalization} 
Recordings were collected over a six-month period in a shared studio, with a notable gap between May and August. This interruption may have introduced inconsistencies in loudness due to variations in recording conditions, such as gain settings or microphone placement. To mitigate potential mismatches between loudness and MIDI velocity across the dataset, we applied a loudness normalization procedure. First, the collected MIDI data were synthesized using \textit{FluidSynth} with a Disklavier-sampled SoundFont\footref{soundfont}. The integrated loudness of each synthesized audio file was measured using the \textit{pyloudnorm} package\cite{steinmetz2021pyloudnorm}. We then computed the average loudness across all rendered files and defined $-$23~\si{LUFS} as the desired global average. A uniform gain offset was calculated based on the difference between the measured average and this global reference. This offset was then applied to each rendered file's loudness to yield a unique target loudness per rendition. Each real PianoVAM audio recording was then scaled to match the target loudness of the corresponding rendition. This process is designed to enhance loudness-velocity consistency while preserving the natural dynamic range of the performances.

\section{Dataset Statistics}\label{sec:dataset-statistics}

The dataset contains 106 solo piano recordings from 10 amateur performers, totaling approximately 21 hours. The repertoire is stylistically diverse, spanning works from 38 composers from the Baroque to modern eras (e.g., Bach, Chopin, Kapustin, Joe Hisaishi) and includes several improvisations. Performers' self-reported skill levels are advanced (70 recordings), intermediate (26), and beginner (10). Although the recording system allowed performers to choose between two performance types ---Performance and DailyPractice--- all recordings were self-labeled as DailyPractice. This indicates that the dataset primarily captures informal practice sessions where strict score adherence cannot always be expected.

To investigate differences in expressive characteristics across datasets, we conducted a brief comparative analysis of MIDI-related distributions between MAESTROv3 and PianoVAM. Specifically, we examined three aspects displayed in \figref{fig:combined-violinplot}: 
\begin{inparaenum}[(i)]
    \item the distribution of pitch (MIDI note numbers), 
    \item the distribution of velocity (MIDI velocity values), and 
    \item the distribution of sustain pedal usage on a per-file basis.
\end{inparaenum}
We computed Cohen's \textit{d} for each musical feature to assess the practical significance of inter-dataset differences.  Pitch (\textit{d} = 0.0446) and velocity (\textit{d} = -0.0379) exhibited negligible differences between datasets. However, pedal usage revealed a large effect size (\textit{d} = 0.870), indicating a substantially higher use of the sustain pedal in PianoVAM compared to MAESTROv3. We speculate this difference stems from an interplay between a pedal-demanding Romantic/Impressionist repertoire, the generous pedaling tendencies of amateur performers in practice, and a less reverberant studio environment that encourages compensation.

\section{Annotation of Fingering Labels}\label{sec:fingering_detection}
\begin{figure}
    \centering
    \includegraphics[width=0.9\linewidth]{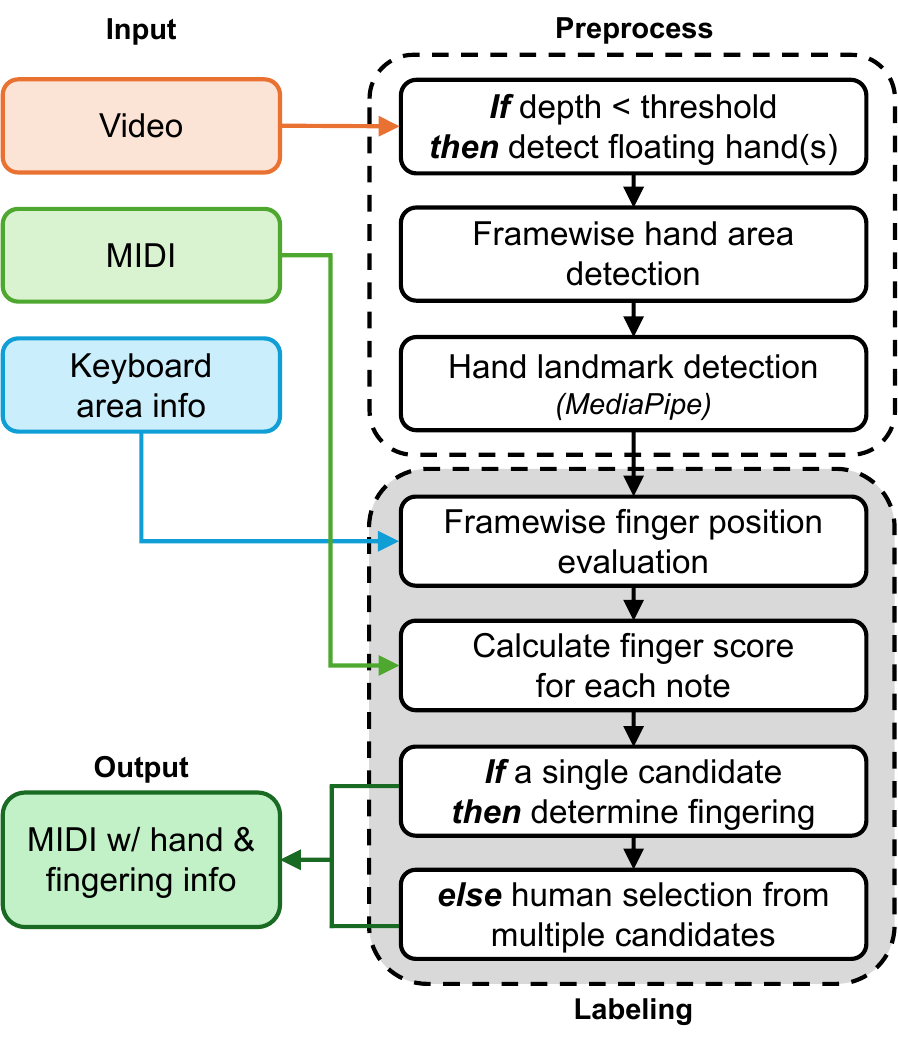}\hspace*{0cm}
    \caption{Flowchart of the fingering detection algorithm.}
    \label{fig:fingering_diagram}
\end{figure}

To generate fingering annotations, we developed the hybrid algorithm shown in \figref{fig:fingering_diagram}. The algorithm first processes performance videos to map hand landmarks to potential finger candidates for each MIDI note. For notes with a single, unambiguous candidate, the fingering is determined automatically, achieving a precision of $\sim$95\% (cf. Table \ref{tab:fingering_results}). In ambiguous cases with multiple candidates (affecting $\sim$20\% of notes), a custom GUI prompts a human annotator to make the final selection. This approach ensures complete and accurate fingering annotations for the entire dataset.

\subsection{Inputs \& Outputs}
The inputs are video, MIDI, keyboard corner locations and lens distortion coefficients in the first frame for each video, which can also be set manually by the GUI-based annotation tool.
The outputs are fingering information, and a separate MIDI file for each hand.

\subsection{Method}

The algorithm suggests candidates fingers that are likely to play each note in the MIDI file. First, hand landmark information is extracted from the input video frame with MediaPipe Hands~\cite{arXiv20Zhang}. A floating hand obfuscating the other hand but not playing any notes should be detected and excluded from the pool of candidates. Thus, we define a metric to measure the $z$-depth of a hand to detect such floating hands. Assuming that we know the hand skeleton of the player explicitly, we can calculate the $z$-depth from the $xy$ coordinate information. For each video, we find the \textit{model skeleton} of each hand, which is a standard for all skeletons that should be unbent, not tilted, and on the keyboard. To measure tilt, the plane defined by three hand key points, namely Wrist ($W$), Index Finger Metacarpals ($I$), and Ring Finger Metacarpals ($R$), is utilized. We assume that the angle $\angle IWR$ of the untilted skeleton (parallel to the plane of the keyboard) should be close to $28\degree$, which is our heuristic estimate for the average human hand in its neutral position. To measure how much the hand is bent, we calculate the ratio $r=\frac{|\triangle IWR|}{|WF_{1}F_{2}F_{3}F_{4}F_{5}|}$ of $|\triangle IWR|$ and area of the hexagon $|WF_{1}F_{2}F_{3}F_{4}F_{5}|$ where $F_i$ is the fingertip of the $i$th finger. Finally, assuming that the hand is playing in the majority of frames, the median $\triangle IWR$ is selected as the default position:
\begin{equation}
    I_{0}W_{0}R_{0}=\underset{|\angle IWR - 28\degree| \text{ low 10\%}, r \text{ high 50\%}}{\text{median}}|\triangle IWR| .
\end{equation}
Let the plane of the projected 2D image be $z=z_{0}$, and the area of 2D image be
\begin{equation}
  \begin{aligned}
    A:=\{(x,y)  | x\in (-1,1), y\in (-AR,AR)\} \subset \mathbb{R}^2
    \\\cong\{(x,y,z_{0})|x\in(-1,1), y\in (-AR,AR)\}\subset \mathbf{P}\mathbb{R}^3 ,
\end{aligned}  
\end{equation}
where the origin of real projective space $\mathbf{P}\mathbb{R}^3$ is the center of the camera lens and $AR$ is the aspect ratio of the video. Since the goal is to calculate the relative depth rather than the real depth, we set $z_{0}=1$ for convenience of calculation.
Knowing that the original point is contained in the line $\{(xt,yt,t) | t\in \mathbb{R^{+}}\}$, we have three equations likely in general position and three variables to solve $t,u,v>0$ from
\begin{gather}
    \left||(x_{I}t,y_{I}t,t),(x_{W}u,y_{W}u,u)|\right|_{\mathbb{R}^3}=\left|\overline{I_{0}W_{0}}\right| \\
    \left||(x_{W}u,y_{W}u,u),(x_{R}v,y_{R}v,v)|\right|_{\mathbb{R}^3}=\left|\overline{W_{0}R_{0}}\right| \\
    \left||(x_{R}v,y_{R}v,v),(x_{I}t,y_{I}t,t)|\right|_{\mathbb{R}^3}=\left|\overline{R_{0}I_{0}}\right|.
\end{gather}

Here, our desired solution can be approximately calculated using Powell's dog leg algorithm with a close initial guess $(t_i,u_i,v_i)=(1,1,1)$. By substituting the solution $(t,u,v)$, we get the 3D coordinates of $I,W,R$.  Finally, $d=(t+u+v)/3$ becomes the relative $z$-depth of the mass of the center of the hand skeleton of each frame. If the $z$-depth is less than the threshold $0.9$, or equivalently, if the hand is floating more than 10\% of the distance between the camera and the keyboard, we decide that the target hand of the target frame is floating.

After discarding all floating hands from the detection, possible finger candidates for each note can be chosen. For each note, a fingering score is calculated, indicating the likelihood of each finger pressing the note. The score is based on the number of frames in which the fingertip is in the selected note area. If the fingertip is completely in the area, a value of 1 is assigned; if it is slightly off, a correction weight is applied, reducing the score for this frame. Thus, the final fingering score cannot exceed the total number of frames of each note. From the fingering score of each finger for the $n$th note, we add the finger as a normal candidate or a strong candidate if the fingering score is greater than 50\% or 80\% of the theoretical maximum score (note length in frames), respectively. If there exists only one strong candidate, we pick the strong candidate as the only candidate. Note that there might be notes with either no candidate or multiple candidates, in which case the algorithm will leave these notes unlabeled.

\subsubsection{Reliability}
With the support of the dedicated GUI-based fingering annotation tool, we manually labeled the ground truth fingering for the first 150 notes of 10 pieces in PianoVAM to assess the results. As detailed in Table \ref{tab:fingering_results}, the algorithm achieved an average precision of over 95\%, with most prediction errors involving adjacent fingers. The table also reports the percentage of notes with no candidates and multiple candidates. Notably, Ravel's \textit{Jeux d'eau} stands out as an outlier with a high ratio of notes without candidates. Excluding this piece, the weighted averages for notes with no and multiple candidates are 9.4\% and 5.2\%, respectively.

\begin{table}
\def\sym#1{\ifmmode^{#1}\else\(^{#1}\)\fi}
\centering
\small

\begin{center}
\begin{tabular*}{\columnwidth}{l@{\extracolsep{\fill}}ccc}
\toprule
\multirow{2}{*}{\textbf{Piece (Index)}} &  \textbf{Prec.} & \textbf{Total}  & \textbf{Total} \\                                     &  \textbf{(150)} & \textbf{No C.} & \textbf{Multi C.}\\
\midrule
Chopin, Op.18 (8) &  91.7 & 12.7 & 8.0 \\
Debussy, L.75 Mvt.3 (17)\sym{*} & 97.1 & 7.9 & 2.8 \\
Grieg, Op.16 Mvt.2 (18) & 99.3 & 10.6 & 4.4 \\
Yiruma, Kiss the Rain (29) & 95.2 & 5.9 & 8.7 \\
Improvisation (31) & 98.6 & 16.7 & 3.0 \\
Schumann, Op.17 Mvt.1 (34) & 82.9 & 8.0 & 5.4 \\
Kapustin, Op.40 No.6 (42) & 98.4 & 8.4 & 5.5 \\
Scarlatti, K.380 (60) & 100.0 & 3.7 & 3.5 \\
Ravel, M.30 (81) & 92.4 & 35.1 & 4.4 \\
Satie, Gymnopedie No.1 (93) & 100.0 & 4.1 & 2.4 \\ 
\midrule
\textbf{Average of 10 pieces} & 95.6 & 13.0$^\dagger$ & 5.1$^\dagger$
\\
\bottomrule
\end{tabular*}
\caption{Precision (\textit{Prec.}) over first 150 notes, percentage of notes with none or multiple candidates (\textit{C.}) over all notes for selected 10 pieces (\sym{*}For finger substitutions, we admit both fingers as correct fingering; $^\dagger$Weighted average with the number of total notes of each piece as the weight).}
\vspace{-7mm}  
\label{tab:fingering_results}
\end{center}
\end{table}

\section{Benchmark Results}\label{sec:transcription}

To demonstrate its utility, we benchmark the PianoVAM dataset on the task of piano transcription under two settings: audio-only and audio-visual. The audio-visual experiments are specifically designed to assess the visual modality's contribution to enhancing performance under challenging acoustic conditions.

\subsection{Data Split}
To facilitate reproducibility of results, we provide information on data splits designed to meet the following criteria: 
\begin{inparaenum}[(i)]
    \item no composition appears in more than one split, and 
    \item the dataset is divided approximately into 80/10/10 percent for the training, validation, and test set, respectively, based on total duration. 
\end{inparaenum}
The resulting train/validation/test splits contain 73, 19, and 14 files, respectively. While these splits are intended to support reproducibility and comparability, we acknowledge that different experimental objectives might require different splits.

\subsection{Audio-Only Piano Transcription}
As MAESTRO is widely regarded as a standard dataset in piano transcription research, we deemed it a suitable reference point for evaluating our dataset. Accordingly, we performed a comparative analysis using the Onsets and Frames model \cite{ISMIR18Hawthorne}, following its original specifications. The model was trained on each dataset as well as on a combined version. We utilized the model weights corresponding to the checkpoint with the lowest validation loss for inference. 

The results in Table~\ref{tab:performance_comparison} are reported as F1 Scores (\%) and calculated over the entire duration of the respective test splits. The terms \textit{Note}, \textit{w/ Offset}, and \textit{w/ Vel.} refer to note evaluation with onset, with onset \& offset, and with onset \& velocity, respectively (cf. \cite{TASLP21Kong}). All four evaluation metrics, including \textit{Frame}, were computed using the \textit{mir\_eval} package\cite{raffel2014mir_eval}. Following established transcription research conventions, offset timings were adjusted to the pedal-release time if the sustain pedal remained engaged.

\begin{table}
\centering
\small
\begin{tabular*}{\columnwidth}{l@{\extracolsep{\fill}}cccc}
\toprule
\textbf{Train Dataset} & \textbf{Note} & \textbf{w/ Offset} & \textbf{w/ Vel.} & \textbf{Frame} \\
\midrule
    MAESTROv3 & 93.4 & 62.3 & 90.3 & 78.2 \\
    PianoVAM & \underline{\textbf{95.8}} & 60.4 & \underline{\underline{\textbf{93.9}}} & 80.0 \\
    Combined & \underline{95.2} & \underline{\underline{\textbf{73.5}}} & \underline{93.0} & \underline{\underline{\textbf{86.9}}} \\
\bottomrule
\end{tabular*}
\caption{Transcription F1 scores on the PianoVAM test split. Bold: highest; Underline: significantly higher than the lowest; Double-line: significantly higher over both others ($p < .0167$).}
\vspace{-4mm} 
\label{tab:performance_comparison}
\end{table}

Statistical tests confirm that significant differences across training sets for all metrics (Friedman test, $p < .001$). Post-hoc Wilcoxon tests with Bonferroni correction ($\alpha = .0167$) showed that both PianoVAM and Combined models significantly outperformed MAESTROv3 in \textit{Note} and \textit{w/ Velocity}. For \textit{w/ Offset} and \textit{Frame}, only the Combined model yielded significantly higher than both others. While PianoVAM slightly outperformed Combined in \textit{Note} and \textit{w/ Velocity}, only the latter difference was statistically significant.

\subsection{Audio-Visual Piano Transcription}

Various approaches have been explored for piano transcription when both audio and video are available. For instance, Wan et al. and Wang et al.\ proposed methods  enhancing the output of an audio-only AMT system by incorporating visual information \cite{CJE15Wan, DAFx21Wang}, while Li et al.\ utilized both modalities jointly to improve onset prediction \cite{ICASSPW23Li, TASLP24Li}. 

For this benchmark experiment, we focus on examining how visual information can be used to improve transcription performance under suboptimal recording conditions. We implement a simple post-processing pipeline that refines MIDI outputs from an audio-only AMT model by using top-view video, estimated piano keyboard corner coordinates, and hand skeletons detected with MediaPipe Hands \cite{arXiv20Zhang}. This process enables the elimination of physically implausible notes by referencing visual evidence, thereby improving onset precision. The full implementation and additional details are available on GitHub\footref{github-link}, and a brief overview follows.

First, onset events are extracted from the predicted MIDI file. For each onset, the nearest video frame is retrieved, and hand landmarks are predicted \cite{arXiv20Zhang}. Each video frame's timestamp is defined as the midpoint of the time interval it covers. If no hand is detected, the corresponding onset is unchanged. When both hands are detected, a perspective transformation is applied using the keyboard corner metadata to produce a normalized rectangular image ($H:W=125:1024$), which maintains the standard height-to-width ratio ($1:8.147$) of an 88-key piano. The same transformation is applied to the predicted hand landmarks. Assuming that the 52 white keys are evenly spaced, the algorithm estimates which white key region each fingertip corresponds to, based on its transformed x-coordinate. To account for possible errors in hand landmark detection, multiple candidate keys are considered for each fingertip, with a tunable threshold determining the candidate range ($\pm2$ white keys in our experiment). The final set of valid pitch candidates is obtained by intersecting all fingertip candidate sets. For each onset, if the pitch predicted by the audio-only AMT model falls within this candidate set, the note is retained; otherwise, it is discarded. This process is repeated for all onsets in the transcription.

\begin{table}
\centering
\small
\begin{tabular*}{\columnwidth}{ll@{\extracolsep{\fill}}ccc}
\toprule
\textbf{Input} & \textbf{Method} & \textbf{Precision} & \textbf{Recall} & \textbf{F1} \\
\midrule
\multirow{3}{*}{Noisy} 
    & Vanilla     & 96.0 & 43.7 & 57.2 \\
    & + NoiseAug  & 96.1 & \textbf{\underline{82.8}} & \underline{88.7} \\
    & + Video     & \textbf{\underline{97.2}} & 82.7 & \textbf{\underline{89.2}} \\
\midrule
\multirow{2}{*}{Reverberant} 
    & Vanilla     & 66.8 & \textbf{68.2} & 64.4 \\
    & + Video     & \textbf{\underline{68.1}} & 67.8 & \textbf{64.8} \\
\bottomrule
\end{tabular*}
\caption{Onset prediction performance under different acoustic conditions. Bold: highest in each column; Underline: significantly higher over the preceding method (paired $t$-test, $p < 0.05$).}
\label{tab:onset_performance_combined}
\end{table}

Table~\ref{tab:onset_performance_combined} summarizes onset prediction performance under two challenging acoustic conditions: SNR=0\si{dB} Gaussian noise, and added reverberation. To evaluate the model's robustness under reverberant acoustic conditions, we applied convolutional reverb using a real-world impulse response (IR) recorded in St.~George's Church\footnote{\href{https://webfiles.york.ac.uk/OPENAIR/IRs/st-georges-episcopal-church/st-georges-episcopal-church.zip}{https://webfiles.york.ac.uk/OPENAIR/IRs/st-georges-episcopal-church/st-georges-episcopal-church.zip}; st\_georges\_far.wav (Last accessed: September 10, 2025)}. The IR was originally sampled at 96\si{kHz} and downsampled to 16\si{kHz} to match the audio input. All audio samples were convolved with the mono version of this IR using FFT-based convolution. To compensate for the inherent delay in the IR (with its peak located at sample index 653), we removed the first 653 samples from each convolved output to ensure proper temporal alignment. The resulting signals were then peak-normalized to maintain consistent amplitude and avoid distortion.

Under noisy conditions, the baseline model (\textit{Vanilla}), trained on clean audio only, exhibited substantial degradation. Introducing noise during training (\textit{+ NoiseAug}) significantly improves recall and F1 ($p < .0001$), while precision remains unchanged. For the \textit{+ NoiseAug} condition, the model was trained on a 50/50 mixture of original clean audio and augmented noisy samples. The noisy samples were generated by adding Gaussian noise with signal-to-noise ratios (SNR) randomly sampled from 0 to 24\si{dB} (cf.~\cite{ISMIR24Kim}). Adding visual filtering (\textit{+ Video}) further improves precision ($p = .0052$) and F1 ($p = .0101$), however, the gain in recall is not statistically significant.

In reverberant conditions, visual post-processing significantly improves precision ($p = .0005$) and marginally improves F1 ($p = .0508$), with no significant change in recall. Qualitative inspection revealed that reverberant tails were sometimes misclassified as new onsets and the visual modality helped reduce such errors.

\section{Discussion}\label{sec:discussion}
The dataset was collected using a system designed to facilitate unsupervised recording, allowing performers to play freely without on-site assistance. While this approach streamlines data acquisition, the dataset exhibits biases in performer identity, pedal usage, and composer representation. In addition, since all recordings originate from practice sessions, the dataset is unsuitable for comparative studies with corresponding musical scores. Our fingering detection approach, while promising, faces challenges from visual ambiguities. These arise from both complex pianistic techniques, such as the multi-finger preparations for rapid repetitions in Chopin's \textit{Grande valse brilliante}, and visual artifacts like performance-induced motion blur in Ravel's \textit{Jeux d'eau} or shadows in the Schumann's \textit{Fantasie in C} recording. Furthermore, the algorithm is intentionally focused on conventional playing, thus excluding extended techniques like glissandi or playing with the fist. Our future extensions may include expert performances, expanded modalities (e.g., multi-angle video), and contextually rich data to support more robust and musically meaningful analysis. Moreover, we aim to improve fingering detection precision by leveraging state-of-the-art models for hand pose estimation \cite{ViTPose++} and 3D reconstruction \cite{dong2024hamba}.

\section{Conclusion}\label{sec:conclusion}
We presented PianoVAM, a comprehensive multimodal dataset of amateur piano practice sessions that captures synchronized top-view video, audio, MIDI, hand landmarks, fingering labels, and rich metadata. Recorded using a Yamaha Disklavier in natural, varied practice conditions, PianoVAM addresses key limitations of existing datasets that often lack specific modalities or rely on synthetic or incomplete annotations. To generate fingering labels, we propose a semi-automated method that combines hand landmark detection from video with manual refinement. We also discuss the challenges of multimodal alignment and data collection. To demonstrate the utility of PianoVAM, we report baseline results for both audio-only and audio-visual piano transcription tasks and showcase its potential for advancing a range of MIR applications. Future extensions of the dataset may address current imbalances in musical content and metadata diversity.

\section{Ethics Statement}
This study involved human participants for data collection, which was approved by the Institutional Review Board (IRB) at KAIST (Approval No. KH2023-235). All procedures strictly adhered to established ethical guidelines.

\section{Acknowledgments}
We sincerely appreciate the KAIST music and audio computing lab and PIAST (Piano club) members who participated in the dataset acquisition as performers. This research was supported by the National Research Foundation of Korea (NRF) funded by the Korea Government (MSIT) under Grant RS-2023-NR077289 and Grant RS-2024-00358448.

\bibliography{ISMIRtemplate}

\end{document}